\documentstyle[preprint,eqsecnum,aps]{revtex}

\begin{document}
\draft
\preprint{HEP/123-qed}
\title{
Dynamical stripe structure in La$_{2-x}$Sr$_x$CuO$_4$ 
observed by Raman scattering 
}

\author{S. Sugai and N. Hayamizu}
\address{Department of Physics, Faculty of Science, 
Nagoya University, Chikusa-ku, Nagoya 464-8602, Japan} 

\date{\today}
\maketitle

\begin{abstract}
	 
The dynamical stripe structure relating to the "1/8 problem" 
was investigated in La$_{2-x}$Sr$_x$CuO$_4$ utilizing the high 
frequency response of Raman scattering.  
The split of the two-magnon peak due to the formation of the 
stripe structure was observed at whole Sr concentration region 
from $x=0.035$ to 0.25 at low temperatures.  
Especially clear split was observed at low carrier 
concentration region $x=0.035 - 0.06$ and at $x \sim 1/8$.  
The onset temperatures of these stripe structures are as high 
as $300-350$ K, which are much higher than the temperatures 
measured by slow response probes.  

\end{abstract}


\narrowtext
	The large exchange interaction energy ($J$) in high 
temperature superconductors contributes to raise the transition 
temperature ($T_{\rm c}$).  
However the large $J$ may induce instability for the uniform 
charge density to separate into spin and charge domains 
\cite{1,2,3,4}.  
Tranquada {\it et al}. \cite{5,6} obtained the experimental evidence 
from the neutron scattering in 
La$_{1.48}$Nd$_{0.4}$Sr$_{0.12}$CuO$_4$ (LNSCO) that the 
antiferromagnetically ordered spin stripes are 
separated by the periodically spacing charge domain 
walls.  
The $T_{\rm c}$ is suppressed in La$_{2-x}$Ba$_x$CuO$_4$ 
(LBCO) and La$_{2-x-y}$Nd$_y$Sr$_x$CuO$_4$ (LNSCO) at 
$x=1/8$ \cite{7,8,9,10}.  
The suppression of $T_{\rm c}$ is related to the structural 
phase transition to the low temperature tetragonal phase 
(LTT, P4$_2$/ncm) below 65 K \cite{11,12}.

	In La$_{2-x}$Sr$_x$CuO$_4$ (LSCO) the decrease of 
$T_{\rm c}$ at $x=1/8$ is only a few degree and the static 
LTT phase is not observed at least down to 4.2 K, but the 
local LTT component increases at low temperatures \cite{8,13,14,15}.    
The magnetic order has been observed below 32 K 
at $x=0.115$ by nuclear magnetic resonance (NMR) and nuclear 
quadrupole resonance (NQR) \cite{16,17}.  
The onset temperature of the wipeout effects for $^{63}$Cu 
NQR decreases monotonically from 90 K at $x=0.07$ to 50 K at 
$x=0.115$ \cite{18}.  
The spatially modulated dynamical spin correlation has been 
observed at $x=0.05-0.25$ as incommensurate magnetic 
peaks by neutron scattering \cite{19,20,21,22,23,24}.  
The direction of the stripe changes from the diagonal 
direction on the CuO$_2$ lattice in the spin glass composition 
($x=0.02-0.05$) to the vertical direction in the superconducting 
composition ($x\ge 0.055$).  
The inter-domain wall distance increases as $l=a/2x$ at 
$x<\frac{1}{8}$ and becomes constant above it, 
where $a$ is the Cu-Cu atomic distance.  
The domain walls are half-occupied by holes at $x<\frac{1}{8}$.  
Above it excess holes enter charge domain walls 
and/or spin stripes.

	The static stripes cause the localization of carriers and 
suppress the superconductivity, so that the dynamical fluctuation 
is essential to induce the metallic conductivity and the superconductivity.  
However, it causes the difficulty for the investigation of stripes, 
because the fluctuating frequency is too high for many 
experimental probes.  
Raman scattering has advantageous for the direct observation 
of high energy magnetic excitations up to near 1 eV.  
Till now intensive Raman scattering studies have been reported in 
LSCO \cite{25,26,27,28,29,30,31,32,33,34}, 
but the evidence of the stripe structure has not been observed.  
In the present Raman scattering experiments, the special attention 
was paid to the sample quality.

	The single crystals were synthesized by the TSFZ method 
utilizing an infrared radiation furnace with four elliptic mirrors 
(Crystal system, FZ-T-4000).  
This method does not use a crucible, so that the crystal is free from 
the contamination.  
The crystal of $x=0$ was annealed in 1 mmHg oxygen gas at 
900$^{\circ}$C for 12 hours.  
The N\'eel temperature ($T_{\rm N}$) measured by SQUID magnetometer 
is 293 K.  
The crystals of $x=0.2$ and 0.25 were annealed in oxygen gas under 
ambient pressure at 600$^{\circ}$C for 7 days.  
The $T_{\rm C}$ was determined by the middle point of the transition 
in the electric resistivity.  
The narrow temperature widths of the superconducting transition region 
(2.5 K at $x=0.115$ and 1.5 K at $x=0.15$) certify the good quality 
of the crystals.  
The details will be presented elsewhere \cite{clpol}.

	Raman spectra were measured on fresh cleaved surfaces in a 
quasi-back scattering configuration utilizing a triple monochromator 
(JASCO, NR-1810), a liquid nitrogen cooled CCD detector (Princeton, 
1100PB), and a 5145 \AA \ Ar-ion laser (Spectra Physics, stabilite 2017).  
The cleavage was done in air and the sample was set in a cryostat 
within ten minutes.  
The cleaved surface presents intrinsic Raman spectra which are 
different from the surface prepared by polishing or chemical 
etching \cite{clpol}.
The laser beam of 10 mW was focused on the area of 50 $\mu$m$\times$500 
$\mu$m.  
The increase of temperature by the laser beam irradiation was less 
than 2 K at 5 K.  
The same spectra were measured four times to remove the cosmic ray 
noise by comparing the intensities at each channel.  
The wide energy spectra covering $12-7000$ cm$^{-1}$ was obtained by 
connecting 17 spectra with narrow energy ranges after correcting the 
spectroscopic efficiency of the optical system.  
The same spot on the surface was measured during the temperature 
variation by correcting the sample position which was monitored by a 
TV camera inside the spectrometer.

	The two-magnon scattering is active in the $B_{\rm 1g}$ 
symmetry and inactive in $B_{\rm 2g}$, if the simple two-magnon 
scattering model is applied to the two-dimensional antiferromagnetic 
(AF) square lattice.  
The $B_{\rm 1g}$ spectrum is obtained in the $(xy)$ 
polarization configuration in which the incident light 
polarization is parallel to $x=[110]$ and the scattered light 
polarization is parallel to $y=[1 \mbox{\=1} 0]$.  
The $B_{\rm 2g}$ spectrum is obtained in the $(ab)$ configuration 
in which the polarizations are parallel to the Cu-O bonds, $a=[100]$ 
and $b=[010]$.

	Figure 1 shows the temperature dependence of the $B_{\rm 1g}$ and 
$B_{\rm 2g}$ Raman spectra.  
In the undoped AF insulating La$_2$CuO$_4$ (LCO), the sharp 
peaks below 700 cm$^{-1}$ are due to one-phonon scattering, the peaks 
from 850 to 1450 cm$^{-1}$ the resonant two-phonon scattering, 
and the peaks from 1700 to 3000 cm$^{-1}$ the four-phonon scattering.  
The peak at 3213 cm$^{-1}$ in the $B_{\rm 1g}$ spectrum at 5 K is 
the two-magnon scattering peak.  
When carriers are doped, the two-magnon peak changes drastically.  
At $x=0.035$, 0.05, 0.06, and 0.115, the two-magnon peaks split into 
double peaks at low temperatures.  
As temperature increases, these split two-magnon peaks decrease in 
intensity and the typical single two-magnon peaks in the coupled 
charge-spin state appear.  
That is, the scattering intensity rises from energy zero to the energy 
corresponding to the temperature, and then increases gradually to the 
two-magnon peak energy, and finally decreases toward over 7000 cm$^{-1}$.  
These high temperature spectra are similar to the spectra in
Bi$_2$Sr$_2$Ca$_{1-x}$Y$_x$Cu$_2$O$_{8+\delta}$ \cite{35}.

	The $B_{\rm 2g}$ two-magnon peak is observed at 4128 cm$^{-1}$ and 
300 K in LCO, which is about 4/3 times the $B_{\rm 1g}$ two-magnon peak 
energy, 3116 cm$^{-1}$ at 300 K.  
It seems that the magnon-magnon interaction does not work in the 
$B_{\rm 2g}$ two-magnon scattering.  
The $B_{\rm 2g}$ spectra are different from the $B_{\rm 1g}$ spectra, 
when the stripe structure disappears at high temperatures.  
Both spectra becomes similar at low temperatures at $x=0.05$, 
0.06, and 0.115.  
It indicates that the symmetry is relaxed in the stripe structure.

	The split two-magnon peaks have fine structure.  
For example at $x=0.06$, the low energy peak is composed of two 
peaks at 1672 cm$^{-1}$ and 2000 cm$^{-1}$ and the high energy 
peak at 2745 cm$^{-1}$ and 3125 cm$^{-1}$.  
It is supposed that these fine structure is induced by the 
interactions between magnons at the nearest neighbor magnetic 
stripes across the charge domain wall, 
because the spin directions are reversed across the charge wall 
and both magnetic stripes give the periodic unit.  
These fine structure will be presented separately.  
In the following the discussions are limited to the double peak 
structure.   
Figure 2 shows the split two-magnon peak energies at 5 K and the 
single two-magnon peak energies at 300 K as a function of $x$.  
The energies of the clearly split two-magnon peaks at $x=0.035$, 0.05, 
0.06, and 0.115 are shown by large circles.  
The intensity of the higher energy peak is much weaker than the 
lower energy peak at $x=0.1$ and $x\ge 0.15$.  
The higher energy peak keeps the energy almost constant, $3050-3350$ 
cm$^{-1}$, from $x=0$ to 0.25, while the lower energy peak decreases 
in energy above $x=0.15$.   
At $x=0.25$ a small peak is observed at 1820 cm$^{-1}$.  
At 300 K where the stripe structure almost disappears, the two-magnon 
peak energy decreases monotonically from $x=0$ to 0.25.

	The origin of the split two-magnon peaks at about 3050 cm$^{-1}$ 
and 1900 cm$^{-1}$ can be explained as follows.  
The two-magnon Raman scattering in the $S=1/2$ antiferromagnet 
is caused by the exchange of nearest neighbor spins.  
Figure 3 shows the case that (a) the spin exchange occurs inside the 
AF square lattice, (b) near the diagonal charge domain wall, 
(c) near the horizontal charge domain wall.  
The case (b) is the same as the two-magnon process in the 
stripe phase of La$_{2-x}$Sr$_x$NiO$_{4+\delta}$ \cite{36,37,38,39}.  
It is supposed that in the charge domain wall the magnetic 
moment at the Cu$^{2+}$ site is zero.  
For the case (a), six bonds shown by the double lines have 
increased exchange interaction energy.  
The total energy increases by about $3J$.  
In the cases (b) and (c) the energy increases at four bonds, 
and then the total energy increases by $2J$.  
Thus the two-magnon scattering energy near the charge domain 
wall is 2/3 times smaller than at the inside of the spin stripe.  
Thus the higher energy peak appears to result from the case (a) 
and the lower energy peak from the case (b) or (c).

	The decrease of the energy of the lower energy peak at 
$x\ge 0.15$ suggests that the carriers are included in the spin 
stripes, although small area of spin stripes has no carrier as 
known from the existence of the higher energy peak at about 3200 
cm$^{-1}$ which is almost the same as or even higher than the 
two-magnon peak energy 3213 cm$^{-1}$ in the undoped AF 
insulating LCO.  
It is consistent with the decreasing Cu-Cu interatomic distance in 
the CuO$_2$ plane, as Sr concentration increases.

	The onset temperature of the appearance of the split peaks 
is obtained from the differential spectra between two temperatures.  
They are plotted in Fig. 4.  
Those at $x=0.2$ and 0.25 should be distinguished 
from other high onset temperatures, because the small sign of the 
higher energy peak is noticeable at 300 K but it does not increase 
at low temperatures.  
For neutron scattering only elastic scattering data are plotted.  
The inelastic incommensurate magnetic peaks have been detected 
from $x=0.06$ to 0.25 at low temperatures, but the temperature 
dependence has not been reported \cite{22}.  
The temperatures determined by Raman scattering are much 
higher than those obtained by other experiments.  
The onset temperature decreases, as the response of the 
experimental probe becomes slow.  
These results indicate that the high frequency fluctuation of 
stripes starts at much higher temperatures than reported.  
The frequency of the fluctuation decreases as temperature 
decreases and the quasi-static component appears at the 
temperatures reported so far.

	In conclusion, the present Raman scattering experiment 
disclosed the existence of the dynamical stripe structure in 
whole carrier concentration region from $x=0.035$ to 0.25.  
These onset temperature is about 300 K for $x=0.035$, 0.05, 
0.06, 350 K for $x=0.115$, 100 K for $x=0.1$ and 0.15.  
These onset temperatures are much higher than those obtained from 
low frequency probes.  
At $x=0.1$ and $x\ge 0.15$ the carrier density in the large 
volume looks uniform, however it may be possible that the 
system is in the isotropic quantum liquid-crystal phase of 
charge strings \cite{40}.
The dynamical fluctuation of stripes is expected to induce 
the metallic conductivity, but it is a continued problem whether 
dynamical stripes help the superconductivity or not.

Acknowledgments - 
The authors would like to thank K. Takenaka for the 
characterization of single crystals and M. Sato for the 
measurement of magnetization.  
This work was supported by CREST of the Japan Science and 
Technology Corporation.

\begin{figure}
\caption[]{(Color)
Temperature dependence of the (a) $B_{\rm 1g}$ and (b) 
$B_{\rm 2g}$ Raman spectra of La$_{2-x}$Sr$_x$CuO$_4$ in 
the antiferromagnetic insulator phase ($x=0$), 
the spin glass phase (below about 10$-$5 K at $x=0.035$ and 
0.05), the superconductor phase ($x\ge 0.06$).  
}
\label{fig1}
\end{figure}

\begin{figure}
\caption[]{
The Sr concentration dependence of the spit two-magnon peak 
energies at 5 K and the single two-magnon peak energies at 
300 K.  
}
\label{fig2}
\end{figure}

\begin{figure}
\caption[]{
The two-magnon Raman process for the case that (a) the spin 
exchange occurs inside the AF square lattice, (b) near the 
diagonal charge domain wall, (c) near the horizontal charge 
domain wall.  
The arrows indicate spins at Cu$^{2+}$ sites.  
The filled circle denotes the presence of one hole.  
When the directions of two neighboring spins are changed by 
$\pm 1$, the exchange interaction energy increases at the double 
lined bonds.  
}
\label{fig3}
\end{figure}

\begin{figure}
\caption[]{
The onset temperatures of the stripe structure, measured by the 
present Raman scattering, EXAFS \cite{14}, the wipeout effect of 
$^{63}$Cu-NQR \cite{18}, $^{139}$La-NMR \cite{17}, and elastic 
neutron scattering \cite{21,23,24}.  
Note that intensities of the split two-magnon peaks at $x=0.1$ 
and $x \ge 0.15$ are weak even at 5 K.
}
\label{fig4}
\end{figure}


\begin{references}
\bibitem{1} K. Machida, 
Physica C {\bf 158}, 192 (1989).
\bibitem{2} D. Poilblanc and  T. M. Rice, 
Phys. Rev. B {\bf 39}, 9749 (1989).
\bibitem{3} J. Zaanen and O. Gunnarsson, 
Phys. Rev. B {\bf 40}, 7391 (1989).
\bibitem{4} V. J. Emery {\it et al}., 
Phys. Rev. Lett. {\bf 64}, 475 (1990).
\bibitem{5} J. M. Tranquada {\it et al}., 
Nature, {\bf 375}, 561 (1995).
\bibitem{6} J. M. Tranquada {\it et al}., 
Phys. Rev. Lett. {\bf 78}, 338 (1997).
\bibitem{7} A. R. Moodenbaugh {\it et al}., 
Phys. Rev. B {\bf 38}, 4596 (1988).
\bibitem{8} K. Kumagai {\it et al}., 
J. Mag. Mag. Materials, {\bf 76}\&{\bf 77}, 601 (1988).
\bibitem{9} Y. Nakamura and S. Uchida, 
Phys. Rev. B {\bf 46}, 5841 (1992). 
\bibitem{10} J. Yamada {\it et al}., 
J. Phys. Soc. Jpn. {\bf 63}, 2314 (1994).
\bibitem{11} J. D. Axe {\it et al}., 
Phys. Rev. Lett. {\bf 62}, 2751 (1989).
\bibitem{12} M. K. Crawford {\it et al}., 
Phys. Rev. B {\bf 44}, 7749 (1991).
\bibitem{13} R. L. Martin, 
Phys. Rev. Lett. {\bf 75}, 744 (1995).
\bibitem{14} A. Bianconi {\it et al}., 
Phys. Rev. Lett. {\bf 76}, 3412 (1996).
\bibitem{15} E. S. Bo\v{z}in {\it et al}., 
Phys. Rev. B {\bf 59}, 4445 (1999).
\bibitem{16} S. Ohsugi {\it et al}., 
J. Phys. Soc. Jpn. {\bf 63}, 2057 (1994).
\bibitem{17} T. Goto {\it et al}., 
J. Phys. Sco. Jpn. {\bf 63}, 3494 (1994).
\bibitem{18} A. W. Hunt {\it et al}., 
Phys. Rev. Lett. {\bf 82}, 4300 (1999).
\bibitem{19} S-W. Cheong {\it et al}., 
Phys. Rev. Lett. {\bf 67}, 1791 (1991).
\bibitem{20} T. R. Thurston {\it et al}., 
Phys. Rev. B {\bf 46}, 9128 (1992).
\bibitem{21} T. Suzuki {\it et al}., 
Phys. Rev. B {\bf 57}, R3229 (1998).
\bibitem{22} K. Yamada {\it et al}., 
Phys. Rev. B {\bf 57}, 6165 (1998).
\bibitem{23} H. kimura {\it et al}., 
Phys. Rev. B {\bf 59}, 6517 (1999).
\bibitem{24} S. Wakimoto {\it et al}., 
Phys. Rev. B {\bf 60}, R769 (1999).
\bibitem{25} K. B. Lyons {\it et al}., 
Phys. Rev. B {\bf 37}, 2353 (1988).
\bibitem{26} S. Sugai {\it et al}., 
Phys. Rev. B {\bf 38}, 6436 (1988).
\bibitem{27} K. B. Lyons {\it et al}., 
Phys. Rev. B {\bf 39}, 9693 (1989).
\bibitem{28} S. Sugai {\it et al}., 
Phys. Rev. B {\bf 42}, 1045 (1990).
\bibitem{29} T. Katsufuji {\it et al}., 
Phys. Rev. B {\bf 48}, 16131 (1993).
\bibitem{30} X. K. Chen {\it et al}., 
Phys. Rev. Lett. {\bf 73}, 3290 (1994).
\bibitem{31} X. K. Chen {\it et al}., 
Physica C {\bf 295}, 80 (1998).
\bibitem{32} J. G. Naeini {\it et al}., 
Phys. Rev. B {\bf 57}, R11077 (1998).
\bibitem{33} J. G. Naeini {\it et al}., 
Phys. Rev. B {\bf 59}, 9642 (1999).
\bibitem{34} J. G. Naeini {\it et al}., 
cond-mat/9909342v2.
\bibitem{clpol} S. Sugai {\it et al}., 
cond-mat/0010174.
\bibitem{35} S. Sugai and T. Hosokawa, 
Phys. Rev. Lett. {\bf 85}, 1112 (2000).
\bibitem{36} G. Blumberg {\it et al}., 
Phys. Rev. Lett. {\bf 80}, 564 (1998).
\bibitem{37} K. Yamamoto {\it et al}., 
Phys. Rev. Lett. {\bf 80}, 1493 (1998).
\bibitem{38} S. Sugai {\it et al}., 
J. Phys. Soc. Jpn. {\bf 67}, 2992 (1998).
\bibitem{39} Blumberg {\it et al}. \cite {36} and Yamamoto 
{\it et al}. \cite{37} assigned two peaks at about 700 cm$^{-1}$ 
and 1100 cm$^{-1}$ to the magnetic origin in 
La$_{1\frac{2}{3}}$Sr$_{\frac{1}{3}}$NiO$_4$.  
Blumberg {\it et al}. concluded that the 700 cm$^{-1}$ 
excitation corresponds to the two spin flip within a single spin 
stripe and the 1100 cm$^{-1}$ excitation to the two spin flip 
at the diagonal sites across the charge domain wall.  
However, we \cite{38} disclosed from the systematic carrier 
concentration dependence in La$_2$NiO$_{4+\delta}$ that the 
700 cm$^{-1}$ excitation is due to a phonon and only the 1100 
cm$^{-1}$ excitation is due to the magnetic excitation related 
to the stripe structure from the following reasons.  
The energy and the temperature dependence of the intensity 
for the 700 cm$^{-1}$ peak are the same as those of the strong 
phonon peak in La$_2$NiO$_{4.00}$.  
The two-phonon peak of this mode is observed at $\delta=0.00$ 
and 0.02.  
The intensity of the 1100 cm$^{-1}$ peak increases instead of 
the decrease of the 1700 cm$^{-1}$ two-magnon peak from the 
carrier-free AF spin region as the carrier concentration 
increases, but the intensity of the 700 cm$^{-1}$ peak is 
independent of the carrier concentration.  
\bibitem{40} S. A. Kivelson {\it et al}., 
Nature {\bf 393}, 550 (1998).
\end{references}
\end{document}